\documentclass{aa}

\usepackage{graphicx}
\usepackage{txfonts}
\usepackage{multirow}

\usepackage[markup=underlined, authormarkuptext=name, authormarkupposition=left]{changes}

\defcitealias{Paper1}{Paper~I}

\newcommand{\tim}[1]{\ensuremath{\times 10^{#1}}}

\def\apec{{\sc apec}}

\def\xspec{{\sc xspec}}

\newcommand{\ergs}{erg\,s$^{-1}$cm$^{-2}$}
\newcommand{\ergPerS}{erg\,s$^{-1}$}

\def\cps{counts s$^{-1}$}
\def\gs{g s$^{-1}$}

\newcommand{\Msun}{$M_{\sun}$}

\newcommand{\SSSJ} {SSS~J122221.7$-$311525}
\newcommand{\SSS} {SSS~J122222}

\begin{document}

\title{Superhumps linked to X-ray emission.}
\subtitle{The superoutbursts of \SSSJ\ and GW~Lib.}
\titlerunning{X-ray observations of \SSSJ.}
   \author{V.~V.~Neustroev\inst{1,2}
          \and
          K.~L.~Page\inst{3}
          \and E.~Kuulkers\inst{4}
          \and J.~P.~Osborne\inst{3}
          \and A.~P.~Beardmore\inst{3}
          \and C.~Knigge\inst{5}
          \and T.~Marsh\inst{6}
          \and V.~F.~Suleimanov\inst{7,8}
          \and S.~V.~Zharikov\inst{9}
          }
   \institute{Finnish Centre for Astronomy with ESO (FINCA), University of Turku, V\"{a}is\"{a}l\"{a}ntie 20, FIN-21500 Piikki\"{o}, Finland\\
              \email{vitaly@neustroev.net}
         \and
             Astronomy research unit, PO Box 3000, FIN-90014 University of Oulu, Finland
         \and
             Department of Physics \& Astronomy, University of Leicester, University Rd, Leicester, LE1 7RH, UK
         \and
             European Space Astronomy Centre (ESA/ESAC), Science Operations Department, 28691 Villanueva de la Ca\~{n}ada, Madrid, Spain
         \and
             School of Physics and Astronomy, University of Southampton, Southampton, SO17 1BJ, UK
         \and
             Department of Physics, University of Warwick, Gibbet Hill Road, Coventry CV4 7AL, UK
         \and
             Institut f\"ur Astronomie und Astrophysik, Universit\"at T\"ubingen, Sand 1, D-72076 T\"ubingen, Germany
         \and
             Kazan (Volga region) Federal University, Kremlevskaya str. 18, Kazan 420008, Russia
         \and
             Instituto de Astronom{\'i}a, Universidad Nacional Aut{\'o}noma de M{\'e}xico, Apdo. Postal 877, Ensenada,
             22800 Baja California, M{\'e}xico
             }

   \date{Received August 4, 2017; accepted ...}


  \abstract
  {We present more than 4 years of {\it Swift} X-ray observations of the 2013 superoutburst, subsequent
   decline and quiescence of the WZ Sge-type dwarf nova \SSSJ\ (\SSS) from 6 days after discovery.}
  {Only a handful of WZ~Sge-type dwarf novae have been observed in X-rays, and until recently GW~Lib was the only
   binary of this type with complete coverage of an X-ray light curve throughout a superoutburst. We
   collected extensive X-ray data of a second such system to understand the extent to which the unexpected
   properties of GW~Lib are common to the WZ~Sge class.}
  {We collected 60 {\it Swift}-XRT observations of \SSS\ between 2013 January 6 and 2013 July 1. Four
   follow-up observations were performed in 2014, 2015, 2016 and 2017. The total exposure time of our
   observations is 86.6 ks. We analysed the X-ray light curve and compared it with the behaviour of
   superhumps which were detected in the optical light curve. We also performed spectral analysis of
   the data. The results were compared with the properties of GW~Lib, for which new X-ray observations
   were also obtained.}
  {\SSS\ was variable and around five times brighter in 0.3--10 keV X-rays during the superoutburst
   than in quiescence, mainly because of a significant strengthening of a high-energy component of the
   X-ray spectrum. The post-outburst decline of the X-ray flux lasted at least
   500~d. The data show no evidence of the expected optically thick boundary
   layer in the system during the outburst. \SSS\ also exhibited a sudden X-ray flux change in the middle
   of the superoutburst, which occurred exactly at the time of the superhump stage transition. A similar
   X-ray behaviour was also detected in GW~Lib. }
  {We show that the X-ray flux exhibits changes at the times of changes in the superhump behaviour of
   both \SSS\ and GW~Lib. This result demonstrates a relationship between the outer disc and the white
   dwarf boundary layer for the first time, and suggests that models for accretion discs in high mass
   ratio accreting binaries are currently incomplete. The very long decline to X-ray quiescence is also
   in strong contrast to the expectation of low viscosity in the disc after outburst.
   }


   \maketitle
%

\section{Introduction}

Accretion discs re-distribute angular momentum to allow stars and planets to form, and generate the
vast amounts of energy which power active galactic nuclei and X-ray binaries. Accretion discs also play
a major role in the overall behaviour of cataclysmic variables (CVs) -- interacting binary
systems in which accretion takes place onto a white dwarf (WD) due to Roche lobe overflow
of a low-mass donor (secondary) component (see review by \citealt{Warner}).

Dwarf novae (DNe) are a subset of CVs with relatively low mass-transfer rates, which undergo outbursts
of 2--8 mag which recur on time-scales of days to years. The outbursts are thought to reflect a thermal
instability of discs which, for certain mass-transfer rates, can never achieve a stable thermal
equilibrium \citep{Meyer81}. In these
cases the disc is either cool and faint, and building up its mass, or it is hot and bright and losing
mass to the WD. For a comprehensive review of the disc instability model (DIM) see \citet{Lasota2001}.
A transition of the disc from the quiescent state to the bright state and back is not an instant process.
The outburst starts at some point in the disc where the thermal instability first occurs and propagates
through the disc in the form of a heating wave switching it from a low-viscosity to a high-viscosity
regime. On the return to quiescence the cooling wave starts near the outer disc and propagates inward
\citep[on transition waves see][]{Meyer84,Cannizzo93}.

DNe of the SU~UMa-type show two types of
outbursts: normal outbursts and less frequent superoutbursts which have a slightly larger amplitude
and a longer duration. A unique property of superoutbursts is the appearance of low-amplitude modulations
with a period of a few percent longer than the orbital one. These modulations are called superhumps and
are explained by a tidal instability of the accretion disc, which grows when the disc expands beyond
the 3:1 resonance radius. This causes the disc to become quasi-elliptical and precess
\citep{Whitehurst88,HiroseOsaki90}.

CVs are well known X-ray emitters. The X-ray emission in non-magnetic CVs is believed to originate from
regions very close to the WD surface \citep{HTCas,ZCha}, usually associated with a boundary layer (BL)
between the accretion disc and the WD \citep{PattersonRaymond2}. The accreting material here
is decelerated from Keplerian to much lower velocities to settle on to the WD surface. This results in
shock heating of the material which is thought to release about half of its accretion luminosity, mostly
in the form of X-rays and EUV emission \citep[for recent reviews see e.g.][]{KuulkersReview,Mukai}.
Though X-ray properties of different DNe are not always consistent, on the whole they can be characterized
by suppression of the X-ray flux during an outburst, accompanied by the softening of the X-ray spectrum
\citep[see, e.g.,][and references therein]{DwarfNovaXrays1,DwarfNovaXrays2}. The evolution of the X-ray
flux of DNe throughout their outburst cycle is thought to be due to changes in the mass accretion rate
with which matter enters the BL, and the corresponding changes in physical conditions in the inner accretion
disc. In quiescence, the mass accretion rate is low, the
BL is optically thin and very hot and assumed to be a source of hard X-rays \citep{PringleSavonije}. In
outburst the mass accretion rate is sufficiently high to produce the optically thick BL, which should
lead to a suppression of the hard X-ray component and a subsequent appearance of the strong EUV emission
\citep{PattersonRaymond2,PattersonRaymond1}.

While this scenario may adequately explain the gross X-ray behaviour of the majority of DNe, there are
still many open issues. For example, several DNe showed an increase of their X-ray luminosity during
an outburst (\object{U~Gem} -- \citealt{UGemXraysHard,UGemXrays}, \object{WZ~Sge} -- \citealt{WZSgeOut},
\object{GW~Lib} -- \citealt{GWLibXrays}, \object{V455~And} -- \citealt{V455And}, \object{ASASSN-15po} --
\citealt{Maccarone2015}). Moreover, during outbursts these CVs showed the appearance
\citep[as in U~Gem:][]{UGemGuver} or a significant strengthening of a high-energy component of the X-ray
spectrum. In general, the nature of the hard X-ray emission during DN outbursts and in quiescence remains
controversial \citep[for a discussion, see][]{Mukai}. In an attempt to explain some of these problems,
a number of different models of the BL have been proposed
\citep[e.g.][]{NarayanPopham93,Medvedev2002,Hertfelder2013}, yet the BL structure is still poorly understood
and even the existence of the BL was put in doubt (the ``missing BL'' problem -- \citealt{MissingBL}).
A disruption of the inner disc was explained e.g. by a weakly magnetized WD \citep{LivioPringle92},
evaporation of the inner disc via a coronal siphon flow \citep{Meyer94}, irradiation from the WD
\citep{King97}.

Interestingly, most of the ``problematic'' DNe mentioned above (GW~Lib, WZ~Sge, V455~And, ASASSN-15po)
belong to an extreme subgroup of the SU~UMa-type DNe, called WZ~Sge-type stars
(\citealt{Vican,WZSgeOpt,MatsuiV455And,Namekata2017}; for a recent review of WZ~Sge stars, 
see \citealt{KatoWZ}). Among the peculiar properties of these objects
are the lack of normal outbursts, a very long duration of superoutbursts (up to a month and even longer),
very large superoutburst amplitudes exceeding 6 mag, very long recurrence time between superoutbursts (years
and decades), and very extended post-outburst declines. It seems that these objects are special because
of the very low mass of the donor star, which supplies matter to the WD with a very low mass-transfer rate
\citep{KatoWZ}. The properties of the WZ~Sge-type stars pose a severe challenge to the DIM, which is
currently unable to explain them \citep{Lasota2001}.

Unfortunately, only a handful of WZ~Sge-type DNe have been observed in X-rays, and, up to the date of
the data presented in this paper, GW~Lib was still the only DN of this type with complete coverage of
an X-ray light curve throughout a superoutburst.
On 2013 January 1, a new WZ~Sge-type DN --- \object{\SSSJ} (hereafter \SSS) --- was discovered in outburst
by the Catalina Real Time Survey (CRTS; \citealt{Drake}). Subsequently \SSS\ was found to have a relatively
bright X-ray counterpart \citep{KuulkersATel} and we therefore started an intensive campaign of ground
and space-based monitoring. An analysis of UV--optical--NIR photometric and spectroscopic observations
of \SSS\ was presented by \citet[hereinafter refered to as \citetalias{Paper1}]{Paper1}.
We showed that the superoutburst with the total amplitude of $\sim$7 mag had an unusual
double structure and a long duration. The two segments of the superoutburst were separated by an $\sim$5
mag and $\sim$10~d dip. The second segment of the superoutburst had a duration of 33 d and it displayed
an extended post-outburst decline lasting $\sim$500 d. The optical light curve of \SSS\ clearly showed
superhumps from the very first time-resolved observations and until at least 420 d after the rapid fading
from the superoutburst. As in all SU~UMa stars, the period of superhumps changed slightly over time
\citep[see also][]{KatoSSS}.

Here, we present the results of X-ray observations of \SSS, which we compare with the X-ray and optical
observations of GW~Lib \citep{GWLibXrays}.
The superoutburst of GW~Lib in 2007 lasted $\sim$26~d and its optical amplitude
exceeded 8 mag \citep{Kato1,Vican}.
Our major result is the discovery of a temporal coincidence of changes in X-ray and superhump behaviour
in both systems, thus linking the properties of the BL with the outer disc.

\begin{table*}
\caption{\label{Tab:SwiftLog}Details of the {\it Swift} observations of \SSS, including count rates.}
\begin{center}
\begin{tabular}{llccllcc}
\hline\hline
HJD mid  &  Obs. ID       &Exp.Time& X-ray count rate &HJD mid  &  Obs. ID     &Exp.Time& X-ray count rate \\
2450000+ &                & (ksec) & (count s$^{-1}$) &2450000+ &              & (ksec) & (count s$^{-1}$) \\
\hline

6299.385 & 00032666001/1  &  1.92 & 0.114~$\pm$~0.008 &6348.749 & 00032666027/1&  0.26 & 0.047~$\pm$~0.007 \\
6299.447 & 00032666001/2  &  2.04 & 0.127~$\pm$~0.008 &6349.397 & 00032666027/2&  0.65 &    (combined)     \\
6301.381 & 00032666002    &  1.98 & 0.082~$\pm$~0.006 &6352.003 & 00032666028  &  1.04 & 0.028~$\pm$~0.005 \\
6303.115 & 00032666003/1  &  1.45 & 0.089~$\pm$~0.008 &6354.805 & 00032666029  &  0.95 & 0.052~$\pm$~0.007 \\
6303.199 & 00032666003/2  &  0.53 & 0.067~$\pm$~0.011 &6357.950 & 00032666030  &  0.94 & 0.045~$\pm$~0.007 \\
6305.329 & 00032666004/1  &  1.80 & 0.071~$\pm$~0.006 &6364.289 & 00032666032  &  0.99 & 0.036~$\pm$~0.006 \\
6305.395 & 00032666004/2  &  0.17 & 0.100~$\pm$~0.025 &6369.639 & 00032666034  &  1.00 & 0.024~$\pm$~0.005 \\
6307.199 & 00032666005    &  1.80 & 0.067~$\pm$~0.006 &6372.710 & 00032666035  &  1.06 & 0.038~$\pm$~0.006 \\
6309.406 & 00032666006/1  &  1.04 & 0.059~$\pm$~0.008 &6376.443 & 00032666036  &  1.11 & 0.029~$\pm$~0.005 \\
6309.472 & 00032666006/2  &  0.91 & 0.042~$\pm$~0.007 &6378.519 & 00032666037  &  1.13 & 0.049~$\pm$~0.007 \\
6310.997 & 00032666007/1  &  1.62 & 0.067~$\pm$~0.006 &6381.599 & 00032666038  &  0.90 & 0.032~$\pm$~0.006 \\
6311.391 & 00032666007/2  &  0.49 & 0.063~$\pm$~0.011 &6384.658 & 00032666039  &  1.02 & 0.024~$\pm$~0.005 \\
6313.198 & 00032666008/1  &  0.87 & 0.060~$\pm$~0.008 &6387.530 & 00032666040  &  0.91 & 0.017~$\pm$~0.004 \\
6313.411 & 00032666008/2  &  1.22 & 0.059~$\pm$~0.007 &6391.320 & 00032666041  &  1.02 & 0.035~$\pm$~0.006 \\
6315.071 & 00032666009/1  &  1.68 & 0.070~$\pm$~0.006 &6394.210 & 00032666042  &  0.94 & 0.027~$\pm$~0.005 \\
6315.133 & 00032666009/2  &  0.55 & 0.081~$\pm$~0.012 &6397.075 & 00032666043  &  0.91 & 0.025~$\pm$~0.005 \\
6317.004 & 00032666010/1  &  0.93 & 0.125~$\pm$~0.012 &6399.962 & 00032666044  &  0.99 & 0.020~$\pm$~0.005 \\
6317.071 & 00032666010/2  &  1.01 & 0.082~$\pm$~0.009 &6407.992 & 00032666046  &  1.03 & 0.039~$\pm$~0.006 \\
6318.944 & 00032666011/1  &  1.32 & 0.033~$\pm$~0.005 &6416.000 & 00032666047  &  1.05 & 0.026~$\pm$~0.005 \\
6319.007 & 00032666011/2  &  0.72 & 0.043~$\pm$~0.008 &6418.949 & 00032666048  &  0.96 & 0.033~$\pm$~0.006 \\
6322.354 & 00032666012    &  1.12 & 0.051~$\pm$~0.007 &6421.980 & 00032666049  &  1.02 & 0.033~$\pm$~0.006 \\
6324.555 & 00032666013    &  1.01 & 0.063~$\pm$~0.008 &6424.929 & 00032666050  &  0.76 & 0.029~$\pm$~0.006 \\
6327.706 & 00032666014    &  1.05 & 0.067~$\pm$~0.008 &6428.338 & 00032666051  &  1.02 & 0.035~$\pm$~0.006 \\
6330.836 & 00032666015    &  1.00 & 0.053~$\pm$~0.007 &6430.733 & 00032666052  &  0.99 & 0.032~$\pm$~0.006 \\
6333.839 & 00032666016    &  1.09 & 0.032~$\pm$~0.005 &6434.017 & 00032666053/1&  0.82 & 0.041~$\pm$~0.006 \\
6337.197 & 00032666017    &  0.28 & 0.020~$\pm$~0.008 &6434.081 & 00032666053/2&  0.19 &    (combined)     \\
6338.122 & 00032666018    &  0.48 & 0.182~$\pm$~0.020 &6436.953 & 00032666054  &  1.09 & 0.017~$\pm$~0.004 \\
6338.385 & 00032666019    &  0.49 & 0.123~$\pm$~0.016 &6439.958 & 00032666055  &  1.06 & 0.027~$\pm$~0.005 \\
6338.647 & 00032666020    &  1.03 & 0.124~$\pm$~0.011 &6442.913 & 00032666056  &  0.86 & 0.027~$\pm$~0.006 \\
6339.385 & 00032666021    &  0.98 & 0.124~$\pm$~0.011 &6445.833 & 00032666057  &  0.86 & 0.024~$\pm$~0.005 \\
6339.918 & 00032666022    &  1.05 & 0.098~$\pm$~0.010 &6449.463 & 00032666058  &  0.46 & 0.031~$\pm$~0.008 \\
6340.450 & 00032666023    &  1.04 & 0.084~$\pm$~0.009 &6451.455 & 00032666059  &  0.73 & 0.022~$\pm$~0.006 \\
6341.850 & 00032666024/1  &  1.17 & 0.084~$\pm$~0.009 &6457.962 & 00032666060  &  1.03 & 0.022~$\pm$~0.005 \\
6341.919 & 00032666024/2  &  1.59 & 0.059~$\pm$~0.006 &6464.171 & 00032666062  &  1.04 & 0.020~$\pm$~0.004 \\
6341.986 & 00032666024/3  &  2.00 & 0.089~$\pm$~0.009 &6467.190 & 00032666063  &  1.04 & 0.028~$\pm$~0.005 \\
6342.045 & 00032666024/4  &  0.20 & 0.056~$\pm$~0.017 &6475.068 & 00032666065  &  1.04 & 0.023~$\pm$~0.005 \\
6342.927 & 00032666025/1  &  0.60 & 0.087~$\pm$~0.012 &6835.086 & 00032666066  &  2.95 & 0.011~$\pm$~0.002 \\
6342.983 & 00032666025/2  &  0.60 & 0.075~$\pm$~0.011 &7038.833 & 00032666068  &  1.91 & 0.014~$\pm$~0.003 \\
6345.652 & 00032666026/1  &  0.21 & 0.066~$\pm$~0.007 &7496.135 & 00032666069  &  2.06 & 0.014~$\pm$~0.003 \\
6345.930 & 00032666026/2  &  0.99 &    (combined)     &7869.791 & 00032666070  &  4.94 & 0.009~$\pm$~0.001 \\

\hline
\end{tabular}
\tablefoot{The Obs. ID is a unique identifier given to every observation taken with {\it Swift}. The Obs. ID
           column also shows the snapshot number after the slash; a snapshot is a time interval spent continuously
           observing the same sky position. The 'X-ray count rate' column indicates the measured X-ray count rate
           of \SSS\ in the energy range 0.3--10 keV during the corresponding snapshot.}
\end{center}

\caption{\label{Tab:SwiftLogGWLib} Log of the {\it Swift} observations of GW~Lib in quiescence in 2017.}
\begin{center}
\begin{tabular}{ccccccccc}
\hline\hline
HJD mid  &  Obs. ID    &Exp.Time& HJD mid  &  Obs. ID    &Exp.Time& HJD mid  &  Obs. ID   &Exp.Time \\
2450000+ &             & (ksec) & 2450000+ &             & (ksec) & 2450000+ &            & (ksec)  \\
\hline

7932.656 & 00030917047  &  2.11 & 7996.466 & 00030917066  &  0.35 & 8007.474 & 00030917080  &  0.55 \\
7980.974 & 00030917048  &  1.15 & 7998.385 & 00030917067  &  0.53 & 8007.819 & 00030917081  &  0.52 \\
7988.806 & 00030917051  &  0.45 & 7999.121 & 00030917068  &  0.48 & 8008.070 & 00030917082  &  0.20 \\
7989.413 & 00030917053  &  0.53 & 7999.380 & 00030917069  &  0.65 & 8008.199 & 00030917083  &  0.62 \\
7989.543 & 00030917054  &  0.41 & 7999.785 & 00030917070  &  0.54 & 8008.470 & 00030917084  &  0.52 \\
7989.938 & 00030917055  &  0.67 & 8000.116 & 00030917071  &  0.52 & 8008.863 & 00030917085  &  0.40 \\
7990.080 & 00030917056  &  0.66 & 8000.501 & 00030917072  &  0.56 & 8009.198 & 00030917086  &  0.71 \\
7990.350 & 00030917057  &  0.60 & 8000.969 & 00030917073  &  0.52 & 8009.334 & 00030917087  &  0.56 \\
7990.750 & 00030917058  &  0.28 & 8001.168 & 00030917074  &  0.52 & 8009.673 & 00030917088  &  0.53 \\
7991.479 & 00030917060  &  0.40 & 8001.440 & 00030917075  &  0.59 & 8010.864 & 00030917089  &  0.54 \\
7991.676 & 00030917061  &  0.54 & 8001.707 & 00030917076  &  0.52 & 8012.800 & 00030917090  &  0.48 \\
7995.053 & 00030917062  &  0.65 & 8002.091 & 00030917077  &  0.56 & 8014.064 & 00030917091  &  0.53 \\
7995.460 & 00030917063  &  0.65 & 8004.316 & 00030917078  &  0.23 &          &              &       \\
7996.267 & 00030917065  &  0.34 & 8007.136 & 00030917079  &  0.52 &          &              &       \\

\hline
\end{tabular}
\end{center}

\end{table*}


\begin{figure*}
\resizebox{\hsize}{!}
{\includegraphics{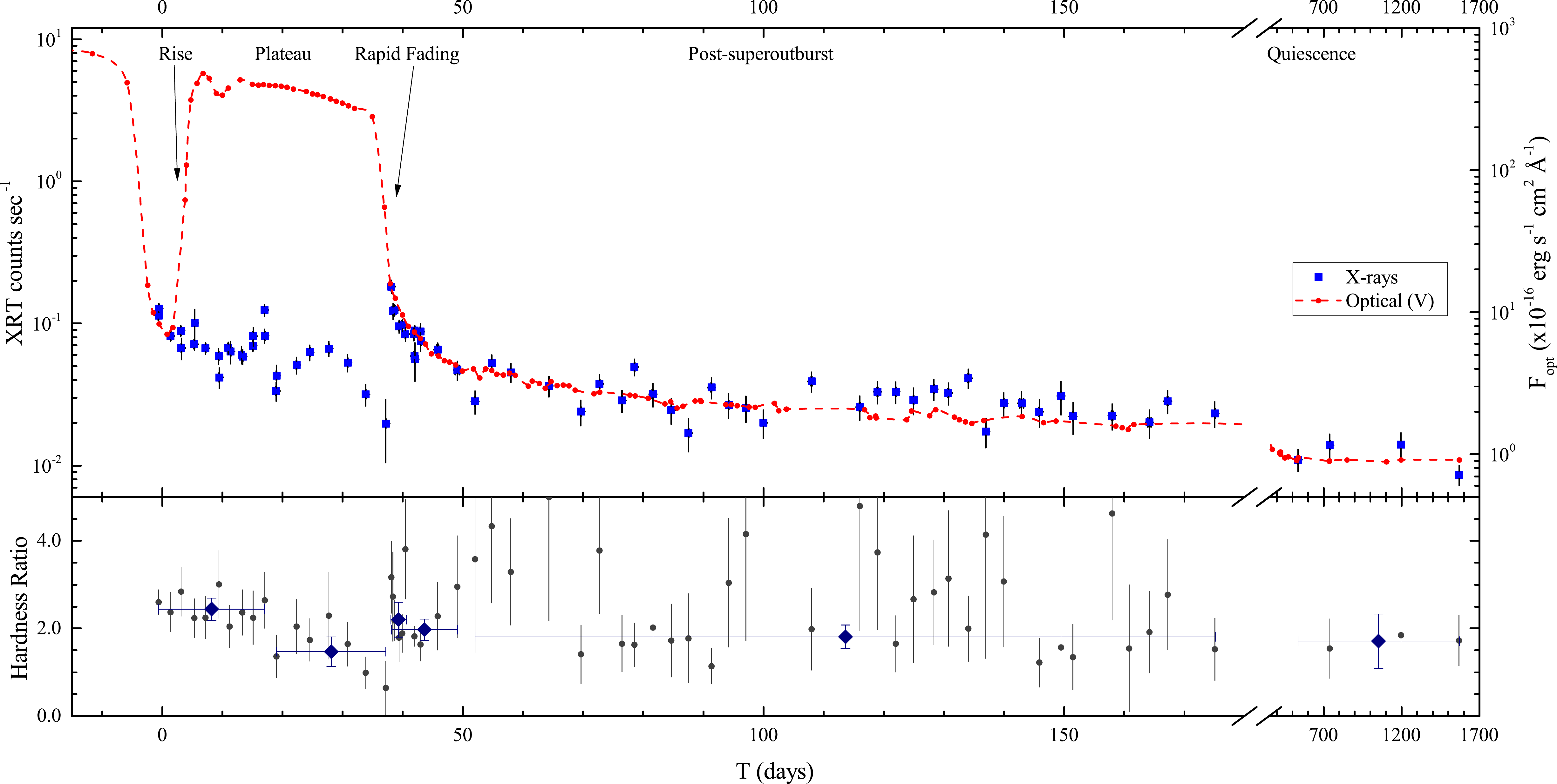}}
\caption{Top: the X-ray light curve of \SSS\ in the energy range 0.3--10 keV shown together with the optical
         $V$-band light curve. $T$ is the number of days elapsed since HJD 245\,6300.0. The optical fluxes
         are converted from 1-d averaged $V$ magnitudes from \citetalias{Paper1}. The left (X-rays) and
         right (optical) axes have the same relative ranges.
         Bottom: the X-ray hardness ratio (1.0--10 / 0.3--1.0 keV) of \SSS. The values averaged over
         representative time ranges are shown by large diamonds.
         }
\label{Fig:LC}
\end{figure*}

\section{Observations and data reduction}
\label{Sec:Obs}

The {\it Swift} X-ray satellite \citep{Swift} started observing \SSS\ on 2013 January 6, 5.8 days after
the discovery announcement \citep{KuulkersATel}. For each observation, data were collected using
both the X-ray Telescope (XRT; \citealt{SwiftXRT}) and the UV/Optical Telescope (UVOT; \citealt{SwiftUVOT}).
In this paper we mostly concentrate on the X-ray observations.

Following the initial exposure, observations were obtained approximately
every 1-3 days until 2013 July 1, with occasional gaps in the schedule caused by high priority {\it Swift}
observations of other targets. A total of 60 observations were taken during this period. Follow-up observations
were performed on 2014 June 26, 2015 January 16 and on 2016 April 17. A final dataset was collected on 2017
April 26. The total exposure time of our observations is 86.6~ks.

The data were processed and analysed using {\sc heasoft} 6.16, together with the most recent
version of the calibration files. All the XRT data were
collected in Photon Counting mode, with the standard grade selection of 0-12 used for the data
analysis. There was no evidence for pile-up at any time, so a circle of radius 20 pixels,
decreasing to 10 pixels as the source faded (1 pixel = 2.36 arcsec), was used for extraction of
the source counts. The background count rate was estimated from a source-free 60 pixel radius
circle, offset from, but close to, \SSS. Table~\ref{Tab:SwiftLog} gives details of the {\it Swift}
XRT observations.

In this paper we also use archival and new {\it Swift} XRT observations of GW~Lib.
The full description and reduction of data obtained in 2007, 2008 and 2009 is presented in
\citet{GWLibXrays}. On 2017 June 28 we performed a new $\sim$2~ks observation of GW~Lib,
and additional 39 observations were taken in 2017 August--September. Thus, a total of
40 observations with the total exposure time of 19.6~ks were obtained in 2017. Details of these 
observations are given in Table~\ref{Tab:SwiftLogGWLib}. They were 
reduced in the same manner as those of \SSS.


\begin{figure}
\resizebox{\hsize}{!}
{\includegraphics[width=8.5cm]{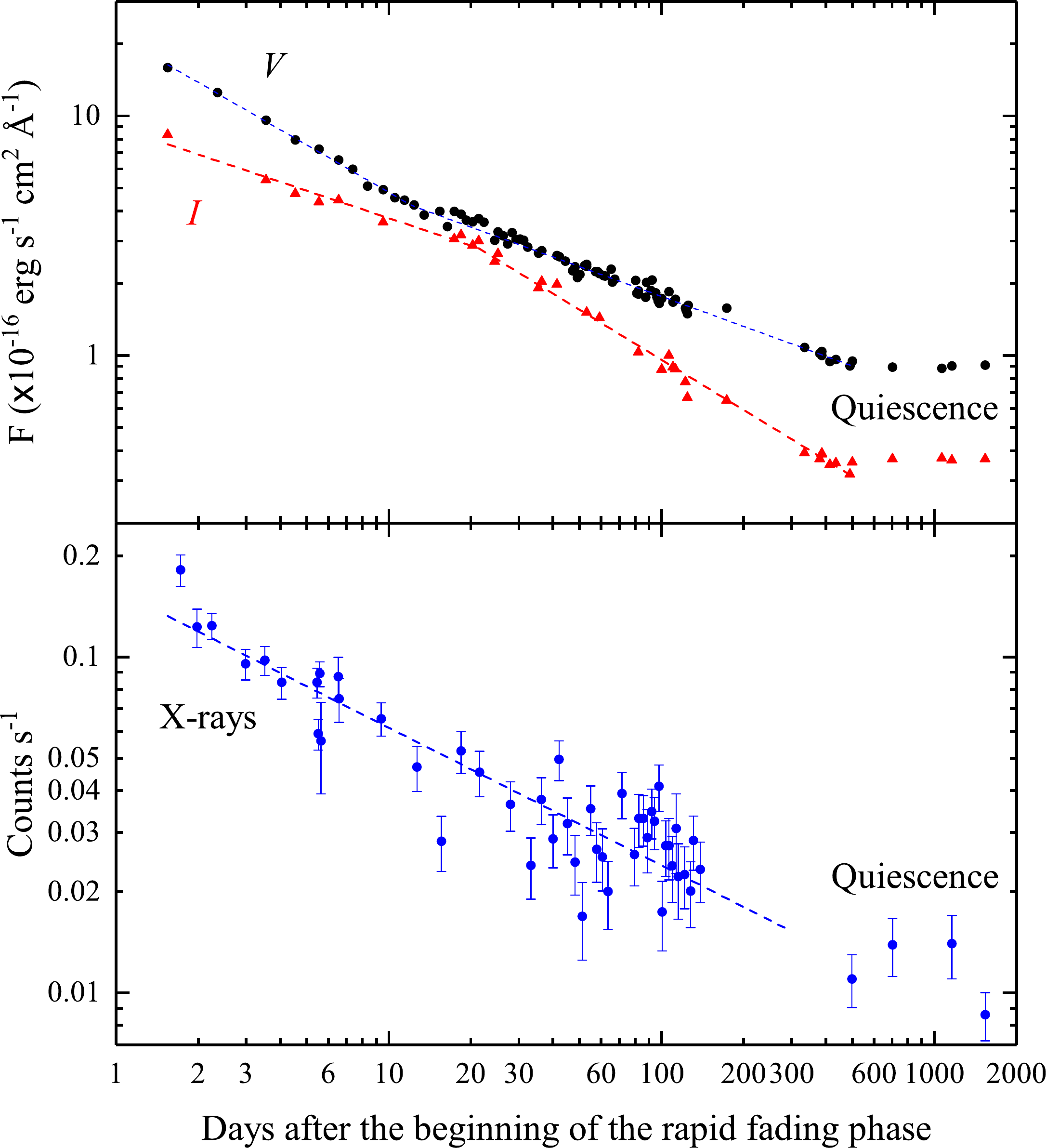}}
\caption{Optical (top panel) and X-rays (bottom) light curves of the decline stage of the
superoutburst of \SSS\ plotted with a logarithmic time scale. The days are counted from
the beginning of the rapid optical fading stage at $T$=36.4.}
\label{Fig:DeclineLog}
\end{figure}

\begin{figure*}
\begin{center}
\hbox{
\includegraphics[height=6.6cm,angle=0]{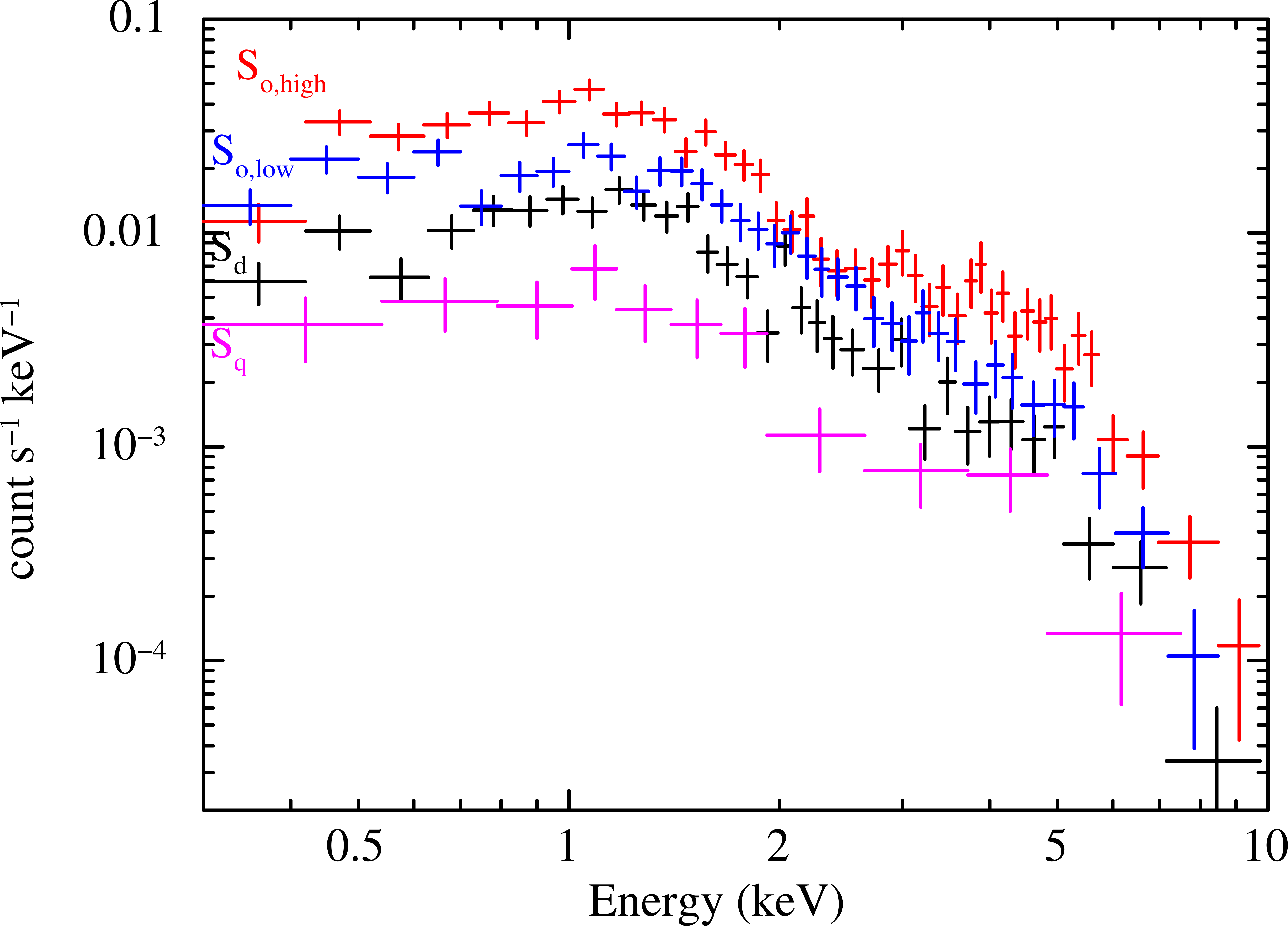}
\includegraphics[height=6.5cm,angle=0]{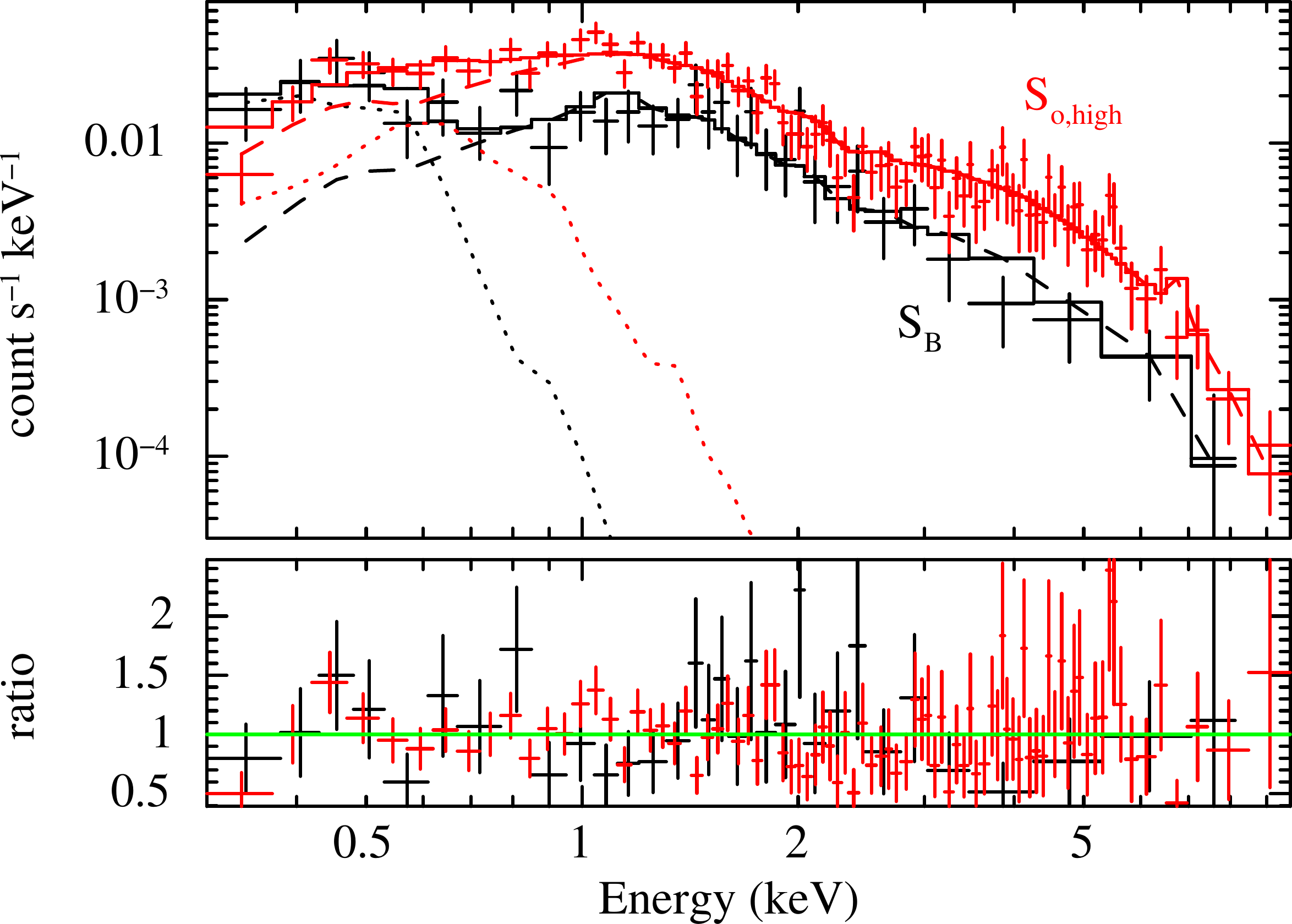}
}
\end{center}
\caption{Left: four X-ray spectra of \SSS\ throughout the superoutburst.
See text for detail.
         Right: Upper panel: best-fit model components of the outburst spectrum S$_{\rm B}$ and the
         brightest outburst spectrum S$_{\rm o,high}$. The data are fitted with two optically-thin thermal
         emission components (shown by dashed and dotted lines), absorbed by  a variable column density. Lower
         panel shows the residuals.}
\label{Fig:Xspecs}
\end{figure*}

\section{X-ray light curve}
\label{Sec:X-rayLC}

Figure~\ref{Fig:LC} (upper panel) shows the X-ray light curve of \SSS\ containing all our X-ray observations
together with the optical fluxes from \citetalias{Paper1}. For consistency, in this
paper we follow \citetalias{Paper1} in defining time $T$ in units of days since 2013 January 8. Thus,
$T$=0 corresponds to HJD 245\,6300.0.
In these units, the first segment of the optical superoutburst began before $T$=$-$22 and the
temporary fading stage began on $T$$\approx$$-$6; the object rebrightened again on $T$=3, reached
the maximum on $T$=7 and faded from the outburst on $T$=37. \SSS\ then displayed an extended post-outburst
decline and returned to quiescence on $T$$\approx$500 (for discussion of properties of the optical light
curve, see \citetalias{Paper1}).

The initial rise of the X-ray and optical flux was missed, and only the second segment of the superoutburst
and following decline had dense coverage in both bands. The X-ray flux of \SSS\ during the superoutburst
plateau ($T$=5--33) was about five times higher than in quiescence ($T$>500): 0.064$\pm$0.022 \cps\ and
0.013$\pm$0.002 \cps, respectively.
The first X-ray observations of \SSS\ were obtained in the middle of the temporary fading stage, just before
the system reached the minimum optical light ($T$=$-$0.6); {\it Swift}-XRT detected a relatively bright
X-ray source with a count-rate of 0.120$\pm$0.006 \cps. On $T$=1.4 the X-ray flux dropped to the
level of about 0.08 \cps\ and then to $\sim$0.07 \cps\ and remained nearly stable until $T$=17, showing
no response to the optical rise to the second segment of the superoutburst. Surprisingly, in the middle
of the outburst the data show a ``zigzag'' --- a strong X-ray peak followed by a drop --- which cannot be
attributed to any feature of the optical light curve: on $T$=17 the count-rate increased sharply to
0.125$\pm$0.012 \cps\ and on $T$=19 was found at 0.033$\pm$0.005 \cps\ (a 7~$\sigma$ difference). During 
the rest of the optical
outburst the X-ray flux showed a slow hump-like evolution around the average level of $\sim$0.05
\cps. On $T$=38.1, at the end of the rapid fading stage, the X-ray light curve exhibited another very
strong peak: during a day the flux rose from 0.02$\pm$0.01 \cps\ to 0.18$\pm$0.02 \cps\ and then
decreased, following the power-law trend which resembles that found for the optical flux.

In \citetalias{Paper1} we found that the decline segment of the optical light curves is well described
by a broken power law as a function of time, counted from $T$=36.4, and
a slope-break some 20~d later. Within the scatter of the data, there is no evidence for a broken power
law in X-ray data (Fig.~\ref{Fig:DeclineLog}). Instead, between $T$$\approx$100--140 the data show a
somewhat higher X-ray flux than expected from a power law with the best fit index $-0.41\pm$0.03 (the
optical data show no such a deviation).

Despite significant flux changes, the hardness ratio (the ratio of counts in the energy bands 1.0--10
and 0.3--1.0 keV) basically remained constant at 1.96$\pm$0.13 from beginning to end of the observations.
However, the hardness ratio data possibly show hints of changes between different superoutburst stages
(Table~\ref{Tab:HR}). The hardness ratio decline before $T$=38 may be step-wise (as shown in Fig.~\ref{Fig:LC},
bottom panel) or a continuous decline, if the former then the hardness ratio difference between
$T$=$-$1 -- 17 and 19 -- 37 is 0.97$\pm$0.42, a 2.3~$\sigma$ difference.

\begin{table}
\caption{The values of the X-ray hardness ratio (1.0--10 / 0.3--1.0 keV) of \SSS, averaged over representative
         time intervals.}
\label{Tab:HR}
\centering
\begin{tabular}{cc}
\hline\hline
 Time interval ($T$)  &  Hardness ratio  \\
\hline
 -1 -- 17    & 2.44$\pm$0.25 \\
 19 -- 37    & 1.47$\pm$0.34 \\
38.1 -- 40.4 & 2.20$\pm$0.40 \\
38.1 -- 49.4 & 1.97$\pm$0.24 \\
 52 -- 175   & 1.81$\pm$0.27 \\
535 -- 1570  & 1.71$\pm$0.62 \\
Entire data set & 1.96$\pm$0.13 \\
\hline
\end{tabular}
\end{table}

There is no evidence for sinusoidal modulation in the X-ray data collected, either during the main outburst
alone or including the decline data. A light-curve was extracted in 5~s bins, and periodograms were calculated
with 256 bins per interval. This placed a 99\% upper limit of 27.6\% on any sinusoidal
modulation (this limit is defined as the amplitude over the mean) for periods between 10 and 640~s.
Thus, the XRT period search does not provide strong constraints on whether or not the WD in \SSS\ is
magnetic given the relatively high modulation amplitude limit. For example, the signature of magnetically
confined accretion is seen in the intermediate polar \object{AE Aqr} which shows an X-ray modulation of
amplitude $\sim$15\% at the WD rotation period of 33~s \citep[e.g.][]{Mauche2006}.


\begin{table*}
\caption{Fits to the X-ray spectra described in the text. Single and double \apec\ models were fitted
to each spectrum. The absorbed and unabsorbed fluxes and unabsorbed luminosities are given in the 0.3--10 keV range.
The luminosity is calculated for the assumed distance of 275 pc for \SSS\ \citepalias{Paper1} and
104 pc for GW~Lib \citep{GWLibXrays}. The final
column gives the probability of improving the fit by including the second temperature component via the F-test.}
\begin{center}
\begin{tabular}{ccccccccc}
\hline\hline
Spectrum   & \apec\ kT1    & \apec\ kT2             & N$_{\rm H}$         & Flux abs.             & Flux unabs.           &Luminosity&C-stat& F-test\\
           &   (keV)       &     (keV)              &(10$^{20}$ cm$^{-2}$)&\multicolumn{2}{c}{(10$^{-12}$ \ergs)}         & \ergPerS &      & prob.\\
\hline
\multirow{2}{*}{S$_{\rm all}$}
\rule{0pt}{4ex}
     & 6.8$^{+1.5}_{-0.8}$ &                        & 2.9$^{+1.4}_{-1.3}$ &                       &                       &            & 538/555\\
\rule{0pt}{2ex}
     &  7.1$^{+1.2}_{-1.1}$       & 0.19$^{+0.08}_{-0.05}$    & 5.5$^{+2.9}_{-2.4}$      &                       &                       &             &532/553 & 95.5\\
\multirow{2}{*}{S$_{\rm o,high}$}
\rule{0pt}{4ex}
     & 8.1$^{+3.1}_{-1.7}$ &                        & 4.1$^{+2.0}_{-2.1}$ & 3.78$^{+0.28}_{-0.25}$& 3.96$^{+0.27}_{-0.24}$&3.58\tim{31}&380/430\\
\rule{0pt}{2ex}
     & 8.5$^{+4.9}_{-2.0}$  & 0.22$^{+0.12}_{-0.06}$ & 9.2$^{+5.8}_{-4.3}$ & 3.82$^{+0.27}_{-0.28}$ & 4.29$^{+0.50}_{-0.36}$&3.88\tim{31}&371/428 & 99.4\\
\multirow{2}{*}{S$_{\rm o,low}$}
\rule{0pt}{4ex}
     & 6.6$^{+4.2}_{-1.3}$ &                        & $<$3.2              & 2.17$^{+0.22}_{-0.16}$& 2.20$^{+0.25}_{-0.15}$&1.99\tim{31}&300/361\\
\rule{0pt}{2ex}
     & 6.5$^{+3.3}_{-1.8}$ & 0.15$^{+0.09}_{-0.06}$ & 5.2$^{+11.8}_{-4.6}$ & 2.20$^{+0.22}_{-0.17}$& 2.39$^{+1.15}_{-0.25}$&2.17\tim{31}&296/359 & 91.0\\

\multirow{2}{*}{S$_{\rm A}$}
\rule{0pt}{4ex}
     & 8.1$^{+3.2}_{-2.4}$ &                        & $<$4.7 & 2.72$\pm$0.26 & 2.72$\pm$0.26 & 2.46\tim{31}&311/342\\
\rule{0pt}{2ex}
     & 6.5$^{+4.2}_{-1.7}$ & 0.14$^{+0.07}_{-0.05}$ & 9.4$^{+11.4}_{-6.5}$ & 2.67$^{+0.31}_{-0.22}$ & 3.12$^{+1.75}_{-0.40}$&2.83\tim{31}&306/340 & 93.6\\
\multirow{2}{*}{S$_{\rm B}$}
\rule{0pt}{4ex}
     & 4.5$^{+2.6}_{-1.2}$ &                        & $<$1.6 & 1.40$^{+0.26}_{-0.27}$ & 1.40$^{+0.36}_{-0.22}$ & 1.26\tim{31}&128/153\\
\rule{0pt}{2ex}
     & 4.2$^{+3.5}_{-1.4}$ & 0.07$^{+0.07}_{-0.02}$ & $<$35 & 1.46$^{+0.27}_{-0.22}$ & 3.7$^{+51.4}_{-2.4}$&3.36\tim{31}&123/151 & 95.1\\

\multirow{2}{*}{S$_{\rm d}$}
\rule{0pt}{4ex}
     & 5.3$^{+1.6}_{-1.1}$ &                        & 5.4$^{+3.0}_{-2.8}$ & 1.12$^{+0.05}_{-0.10}$& 1.20$^{+0.11}_{-0.09}$&1.08\tim{31}&291/322\\
\rule{0pt}{2ex}
     & 5.3$^{+2.0}_{-1.1}$ &  unconst.              & 5.8$^{+4.4}_{-2.7}$ & 1.11$^{+0.11}_{-0.09}$& 1.19$^{+0.13}_{-0.08}$&1.08\tim{31}&290/320 & 42.4\\
\rule{0pt}{4ex}
S$_{\rm q}$
     &  $>$3.7             &                        & $<$9.3              & 0.43$^{+0.13}_{-0.09}$& 0.44$^{+0.12}_{-0.09}$             & 4.01\tim{30}&71/102\\
\\
\hline
\rule{0pt}{4ex}
S$_{\rm GW,q}$
     & 3.5$^{+1.5}_{-0.8}$ &                        & $<$8.9              & 0.50$^{+0.08}_{-0.07}$& 0.53$^{+0.08}_{-0.07}$             & 6.91\tim{29}&135/193\\

\\
\hline
\end{tabular}
\label{Tab:Xfit}
\end{center}
\end{table*}

\section{X-ray spectral analysis}
\label{Sec:XraySpec}

The X-ray light-curve (Fig.~\ref{Fig:LC}) revealed a variable X-ray source. To investigate possible
spectral evolution, seven spectra were extracted from the observations. The spectrum S$_{\rm all}$ consists
of the complete set of XRT data. During the superoutburst stage the X-ray flux varied over a significant
range, roughly between 0.03 and 0.18 \cps. We therefore extracted two spectra for the outburst period.
S$_{\rm o,high}$ consists of the data with the highest XRT count rates $>$0.08 \cps; this covers intermittent
intervals between $T$=$-$1--43. S$_{\rm o,low}$, on the other hand, consists of the faintest outburst
data ($<$0.08 \cps), obtained between $T$=1--37. Bearing in mind that the hardness ratio data show a hint
of a step-wise change at $T$$\sim$18, we extracted two additional spectra, S$_{\rm A}$ and S$_{\rm B}$,
which cover intervals between $T$=$-$1--17 and 19--37, respectively (for the stages A and B of the superoutburst see
Section~\ref{Sec:Xsuperhumps}). The last two spectra, S$_{\rm d}$ and S$_{\rm q}$, consist
of the data collected during the decline stage ($T$=46--175) and after the system had reached the quiescent
level (that is, the last four X-ray observations, $T$=535, 738, 1196 and 1569). The spectra S$_{\rm o,high}$,
S$_{\rm o,low}$, S$_{\rm d}$ and S$_{\rm q}$ are shown in Fig.~\ref{Fig:Xspecs} (left-hand panel).

All the spectra, except for S$_{\rm q}$, were initially fitted with first one, then two, optically thin
emission components (the \apec\ model in \xspec), absorbed by a variable column, N$_{\rm H}$. The Wilms
Solar abundances \citep{Wilms} and Verner photoelectric
absorption cross-sections \citep{Verner} were assumed when fitting with the {\sc tbabs} absorption model.
The final spectrum, S$_{\rm q}$, consists of only 121 counts; therefore, no attempt was made to fit these
data with multiple emission components.

None of the spectra shows a significant difference in the primary \apec\ temperature, with the emission
being around 5--10 keV (see Table~\ref{Tab:Xfit}). Inclusion of a second optically thin emission model
resulted in a better fit (99.4\% confidence) only for the spectrum S$_{\rm o,high}$, for which a temperature
of kT$\sim$0.2 keV was derived. The fit to this spectrum was further improved when we replaced the cooler
\apec\ model with a blackbody model again having kT$\sim$0.2 keV. We consider that these multi-component
fits point to a multi-temperature emission spectrum, for which we do not have sufficient counts to derive
a physically reasonable detailed description. In particular, we do not associate the fitted blackbody
component with an optically thick BL due to its high temperature and low luminosity. The full set of
fitted parameters is given in Table~\ref{Tab:Xfit}. The data and the best-fit model components for
S$_{\rm o,high}$ and S$_{\rm B}$ with residuals are plotted in Fig.~\ref{Fig:Xspecs} (right-hand panel).
In addition, we also extracted the spectrum from the 2017 observations of GW~Lib. This spectrum S$_{\rm GW,q}$
looks very similar to S$_{\rm q}$, so we modelled it in a similar manner. The results are also given in
Table~\ref{Tab:Xfit}.

If a BL between the accretion disc and the WD is present in this system, cool blackbody emission would
be expected with kT $\sim$ a few tens of eV and whose luminosity is expected to be comparable to the
disc luminosity (\citealt{PringleSavonije,PophamNarayan}; such a component
was indeed observed in a few DNe, see e.g. \citealt{Long96,MaucheRaymond,Mauche2004}). Although the XRT
is only sensitive down to energies of 0.3 keV, limits can be placed on the luminosity of such a soft
X-ray blackbody. The S$_{\rm o,high}$ spectrum was used, with a single \apec\ component to model the
major component of the spectrum. A blackbody model was added at temperatures between 10 and 50 eV, with
N$_{\rm H}$ set to 1.6~$\times$~10$^{20}$ cm$^{-2}$ and at 4.4~$\times$~10$^{20}$ cm$^{-2}$ (limits taken
from 90\% confidence range for the best fit to the S$_{\rm all}$ spectrum as detailed in Table~\ref{Tab:Xfit}).
Taking the distance to be 275~pc \citepalias{Paper1}, the 90\% upper limit on the luminosity of a blackbody
was thus estimated to be between $\sim$~10$^{30}$ erg~s$^{-1}$ (for the hottest blackbody temperatures
of $\sim$50 eV) up to $\sim$10$^{37-38}$ erg~s$^{-1}$  (for a very low blackbody temperature of 10 eV).
For a typical\footnote{A preliminary modelling of the UV-optical spectral energy distribution of \SSS\
during the superoutburst gives $L$=2.4\tim{34} \ergPerS.} disc luminosity of a few~$\times$~10$^{34}$
erg~s$^{-1}$, temperatures hotter than $\sim$15~eV would imply only a small fraction of this luminosity
is being emitted in an optically thick BL. At this luminosity the observed UVOT $uvm2$
magnitude of 10--11 would be exceeded by a blackbody BL cooler than kT$\sim$15 eV. Thus, although we
have not seen direct evidence of an optically thick BL, our data do not exclude a kT$\sim$15~eV BL
at the expected luminosity.

\section{Comparison with GW~Lib }

\begin{figure}
\resizebox{\hsize}{!}
{\includegraphics{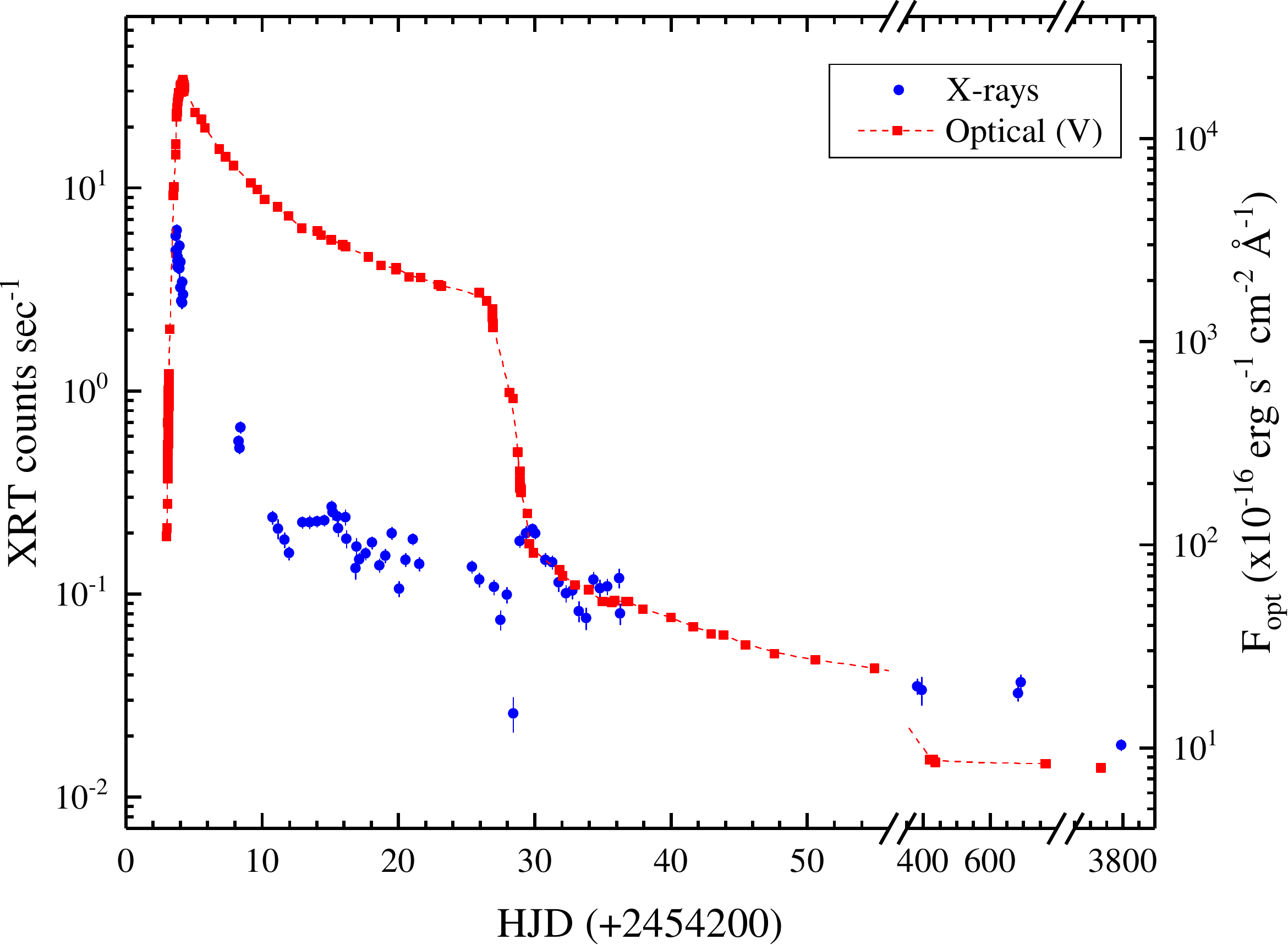}}
\caption{The X-ray light curve of GW~Lib in the energy range 0.3--10 keV shown together with the optical $V$-band light curve.
         The optical fluxes are converted from 1-d averaged $V$ magnitudes from \citet{Kato1} and the American Association of
         Variable Star Observers (AAVSO) data base. The left (X-rays) and right (optical) axes have the same relative ranges.}
\label{Fig:GW_Lib_LC}
\end{figure}

It is instructive to compare the most reliable X-ray properties of \SSS\ with those of other WZ~Sge-type
stars. Unfortunately, of this class, only GW~Lib has been observed in detail, allowing for a comparison.
In contrast to \SSS, the optical light curve of GW~Lib showed a rather usual pattern of superoutburst
with no sign of a dip during the plateau stage (Fig.~\ref{Fig:GW_Lib_LC}). However, the X-ray light
curve displayed a strong peak\footnote{If such a peak was also present in \SSS, we likely missed it.}
around the time of the optical peak with an XRT count rate of $~\sim$5-6 \cps, after that the X-ray flux
declined rapidly for $\sim$10 days before stabilizing at an average level of 0.20$\pm$0.04 \cps\ during
the remainder of the optical outburst. After a short deep dip and a bump observed immediately after the
rapid fading from the outburst, the X-ray flux declined to the level of 0.10$\pm$0.02 \cps. One and two
years after the superoutburst, the X-ray flux was found to be $\sim$0.03$\pm$0.01 \cps\ (for discussion
of properties of the X-ray and optical light curves see \citealt{GWLibXrays} and \citealt{Vican},
respectively). The most recent observations, taken in 2017, showed that the X-ray flux dropped further
to the level of 0.016$\pm$0.001 \cps. We point out, however, that 2 years \emph{before} the superoutburst
the X-ray flux of GW~Lib was about five times below the 2017 level \citep{GWLibXbefore}. Thus, 10 years
after the superoutburst, the object has not yet returned to its pre-outburst level. In contrast to GW~Lib,
the observed quiescent X-ray flux of the prototype system WZ~Sge showed similar values before and 10 years
after its most recent 2001 superoutburst \citep{WZSgeQ}.


%

%
%

It is particularly interesting that in GW~Lib the hardness ratio gradually increased during the superoutburst
plateau from $\sim$1 to $\sim$2 \citep{GWLibXrays}, whereas \SSS\ rather showed a gradual decrease of
the hardness ratio from $\sim$2.5 to $\sim$1. In quiescence the hardness ratio is nearly the same in both
objects: 1.97$^{+0.17}_{-0.20}$ in GW~Lib and 1.71$\pm$0.62 in \SSS, and their spectra are
statistically indistinguishable.
We note that in the very beginning of the superoutburst, the GW~Lib spectrum showed a strong very soft
X-ray emission component consistent with an optically thick BL, which turned out to be undetectable in
later observations \citep{GWLibXrays}. In \SSS\ no supersoft component was ever seen, and the hardness
ratio evolution in this object was probably because of changes in contribution of hard and softer spectral
optically thin components.

%

\section{On the long and short-term variability of the X-ray emission}

The X-ray light curve of \SSS\ shows two
substantial short-term flux changes during the superoutburst stage: 1) the notable zigzag in the
middle of the plateau stage, which had no counterpart in either optical or UV light curves; 2) the very
strong spike observed immediately after the rapid fading. Another unusual property of \SSS\ was an extended
post-outburst decline of the X-ray flux. None of these features is common in DNe,
we discuss them in the following subsections.

\subsection{Transition of the accretion disc from optically thick to optically thin regimes}
\label{Sec:DiscTransit}

The appearance of the spike of X-ray emission after the rapid fading from the superoutburst plateau
of \SSS\ ($T$$\approx$38) is not completely unique. An increase of the X-ray flux after an outburst,
in the form of a `bump', was firstly reported for SS~Cyg \citep{SS_Cyg1,SS_Cyg2} and later for GW~Lib
\citep{GWLibXrays}. These bumps are usually attributed to the transition of the BL from optically thick
to optically thin regimes. \citet{Schreiber03} showed that an increased X-ray flux is expected
at the onset of an \textit{every} outburst when the mass accretion rate is rising and while the BL is
still in an optically thin state, and at its end, when the mass accretion rate decreases below a critical
value of the order of 10$^{16}$ \gs\ and the BL transitions to its optically thin state. Surprisingly,
however, a temporary increase at both the beginning and end of outbursts was observed previously only
in SS~Cyg and GW~Lib \citep{SS_Cyg1,SS_Cyg2,GWLibXrays}.

Although we have no direct spectral evidence that the X-ray peak at the end of the superoutburst of \SSS\
is due to the optically thick/thin state transition, variations of the hardness ratio show a hint of spectral
hardening immediately after the outburst, supporting this interpretation (Fig.~\ref{Fig:LC}). It encouraged
us to attempt to measure the critical mass accretion rate $\dot{M}_{\rm cr}$ of the BL transitioning to its
optically thin state. Following \citet{Fertig}, we assume that the highest luminosity observed in the
beginning of the decline corresponds to $\dot{M}_{\rm cr}$. We derived this luminosity using the observation
obtained on $T$=38.122 (count rate 0.18$\pm$0.02). For a double \apec\ model all the parameters were
frozen at the values listed in Table~\ref{Tab:Xfit} for the S$_{\rm o,high}$ spectrum and only the normalization
was left to vary. We obtained the luminosity in the 0.3--10 keV energy range to be 4.82\tim{31} \ergPerS,
and the bolometric luminosity $L$=9.32\tim{31} \ergPerS. We then converted $L$ to the mass
accretion rate using $\dot{M}$=2$LR_{\rm wd}$/$GM_{\rm wd}$, where we adopted $M_{\rm wd}$=0.9~\Msun\
and $R_{\rm wd}$=6.0\tim{8}~cm \citepalias{Paper1}. As a result, we found $\dot{M}_{\rm cr}$=9.4\tim{14}~\gs.
This value is much lower than the theoretical values \citep{PophamNarayan}, but
is in agreement with \citet{Fertig}, who also derived $\dot{M}_{\rm cr}$ for six DNe and came to the
same conclusion. In this respect, we note that the results of calculations of \citet{PophamNarayan} are
questionable because they underestimated the Rosseland ``true'' opacity by at least two orders of magnitude
\citep{Suleimanov2014}. The use of a more realistic coefficient will decrease the value of
$\dot{M}_{\rm cr}$.

\begin{figure*}
\includegraphics[width=8.8cm]{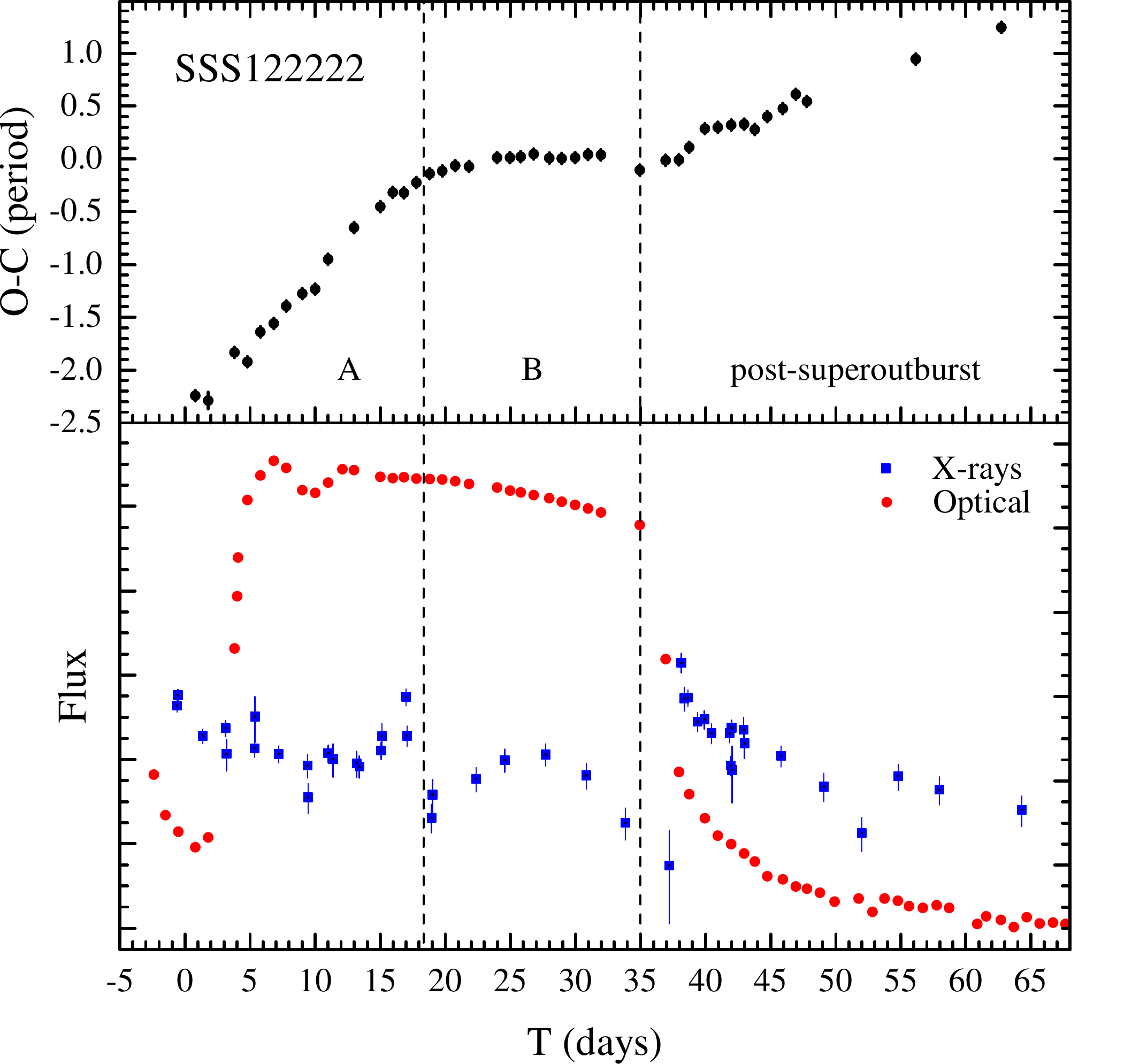}
\hspace{5 mm}
\includegraphics[width=8.8cm]{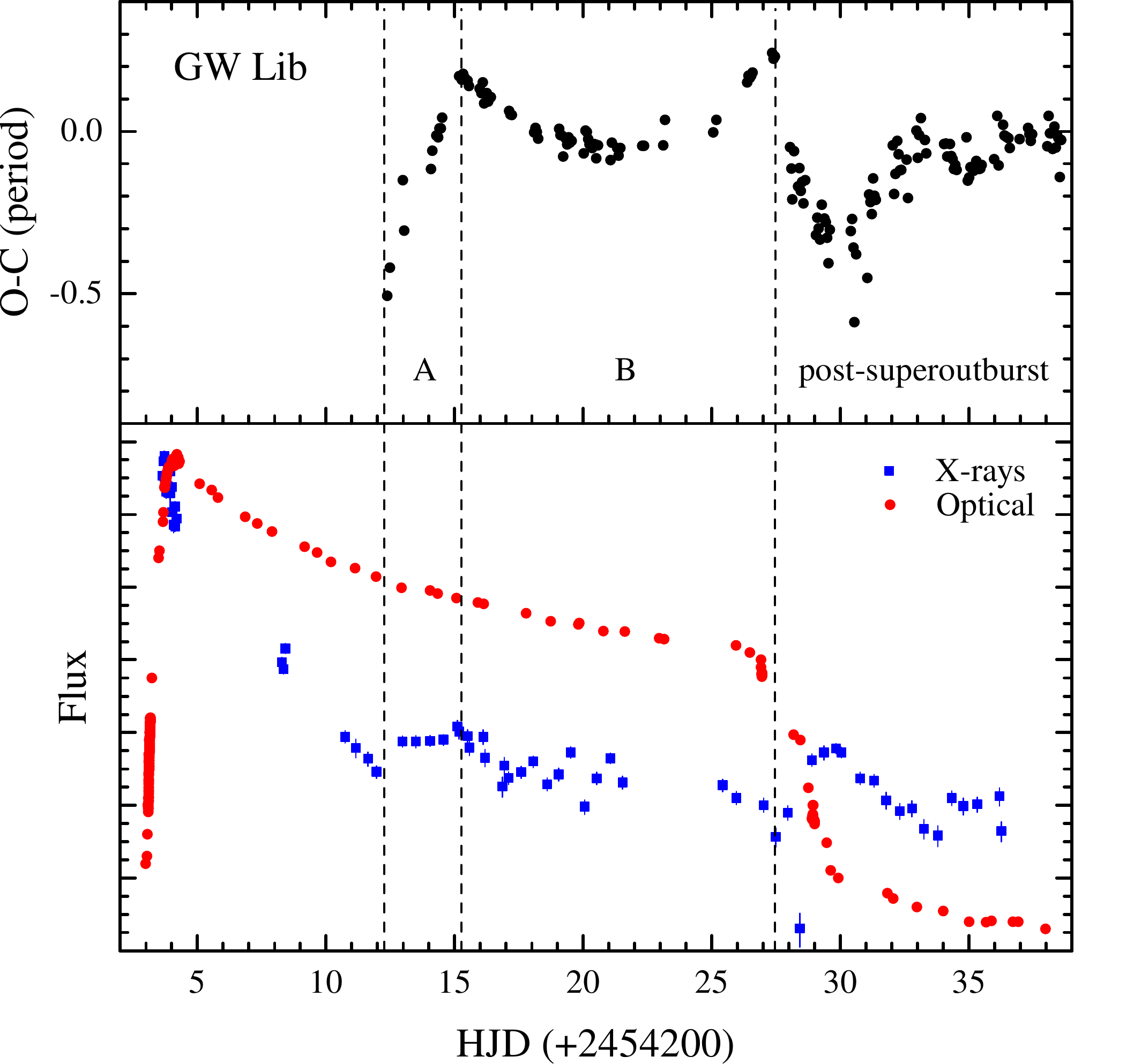}
\caption{$O-C$ diagrams (upper panels) and optical and X-ray light curves (bottom panels, arbitrary
         scaling to emphasise the change in flux) of \SSS\ (left-hand panel) and GW~Lib (right-hand panel).
         The $O-C$ diagram of \SSS\ is taken from \citetalias{Paper1}. To create the $O-C$ diagram of GW~Lib,
         we used the timings of superhump maxima from table 152 in \citet{Kato1}. The stages of the superhump
         evolution are marked following \citet{KatoSSS} and \citet{Kato1}.
         }
\label{Fig:O-C}
\end{figure*}

\subsection{The X-ray zigzag in the middle of an outburst -- a sign of an accretion disc transformation}
\label{Sec:Xsuperhumps}

As described above, the X-ray emission of non-magnetic CVs is believed to arise from close to the WD and
thus reflects the physical conditions in the innermost parts of the accretion disc. A change of these
conditions during a quiescence–outburst–quiescence cycle should lead to an increase or a depression of
the X-ray flux. This scenario, however,
fails to give an explanation of the X-ray zigzag in the middle of the superoutburst of \SSS.
Indeed, in the middle of the outburst the accretion disc is in an almost steady state, the mass accretion
rate is persistently high and the BL is optically thick. Under these conditions no significant variations
in X-ray flux are expected.

In this respect, we point out that during the \textit{superoutbursts} the accretion disc structure might
be more complex than that predicted by the standard disc theory. Indeed, it is known that although the
optical light curves of \SSS\ and other SU~UMa-type stars during the plateau stage of superoutbursts
are very smooth, their accretion discs experience structural changes. An extensive study of superhumps
showed \citep[see][and their later papers]{Kato1} that the evolution of superhump periods in most of
SU~UMa systems is composed of three distinct stages (A, B, and C in \citealt{Kato1}).
 \citet{Kato1} pointed out that the
transitions between the stages are usually fast and abrupt, with discontinuous period changes, suggesting
that these transitions occur due to geometrical and dynamical transformations in the accretion disc.
As a textbook example of such an evolution, Fig.~\ref{Fig:O-C} (upper right-hand panel) shows the $O-C$
diagram of GW~Lib, in which these three stages were observed during the 2007 superoutburst.

Although the superhump period in \SSS\ evolved rather differently from other SU~UMa DNe \citep[for
a discussion of possible causes, see][]{KatoSSS}, nevertheless its $O-C$ diagram (Fig.~\ref{Fig:O-C},
left-hand panel) also shows three distinct segments. Surprisingly,
by comparing the $O-C$ diagram with the X-ray light curve, one can clearly see that the X-ray zigzag in
the middle of the superoutburst occurred during the transition from stage A to stage B. Moreover,
the hardness ratio variations also show a hint of spectral softening just after the zigzag (Fig.~\ref{Fig:LC}).
It is also particularly interesting that the colour profiles of superhumps, which were observed during
these three stages of superhump evolution, were notably different \citepalias{Paper1}.
With such a notable X-ray response, it is tempting to speculate that the accretion disc transformation
may involve the entire disc, from the outer regions where the superhump light arises down to
regions which produce X-rays.

\begin{figure*}
\begin{center}
\includegraphics[width=16cm,angle=0]{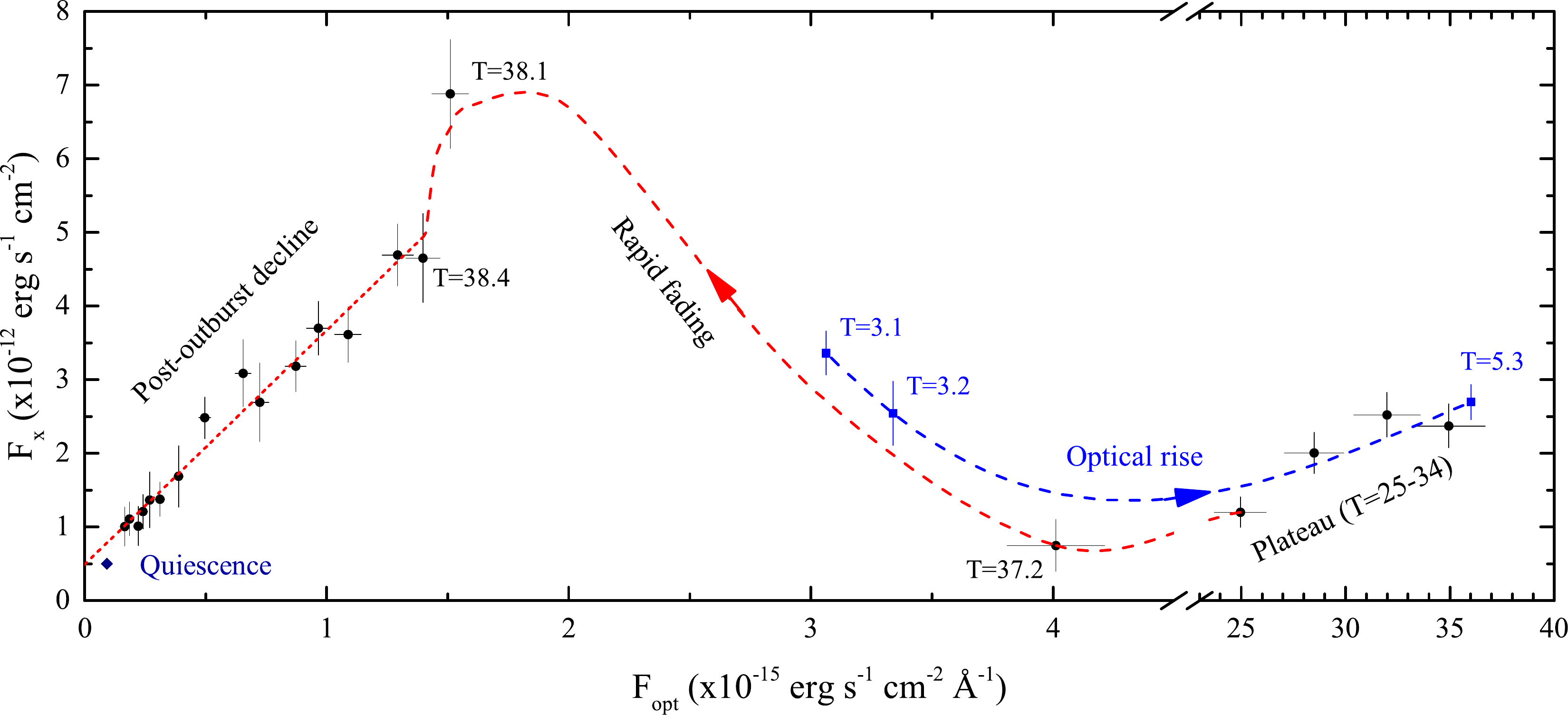}
\end{center}
\caption{The optical/X-ray flux diagram showing the post-superoutburst data and some representative
         data-points from the optical rise (shown in blue) and late plateau stages of \SSS.
         The dashed lines show the assumed trends of flux changes during the optical rise (blue) and
         rapid fading (red). The red short-dashed line shows the linear fit to the data from the
         post-outburst decline stage. The brightest X-ray data-point ($T$=38.1) and the quiescence
         fluxes are seen to lie off the linear dependence.}
\label{Fig:X_Opt}
\end{figure*}

If this was not just a coincidence, then one can expect to observe similar X-ray signatures of the
superoutburst stage transitions in other SU~UMa-type systems. The only other SU~UMa-type CV whose
superoutburst was observed in detail in both the X-rays and in the optical bands and whose superhumps
were studied for period variations is GW~Lib \citep{GWLibXrays,Kato1,Vican}. A comparison of its $O-C$
diagram with the X-ray and optical light curves (Fig.~\ref{Fig:O-C}, right-hand panel) shows several
interesting features. First, the stabilizing of the X-ray flux after the initial rise and the following
rapid decline has occurred simultaneously with the \textit{first} appearance of superhumps about 10 days
after the beginning of the superoutburst. This so-called superhump delay is explained as the growth time
of the tidal instability in the accretion disc \citep{Lubow1}. The rapid decline of the X-ray flux observed
during the first 10~d of the superoutburst indicates that while an eccentric disc was developing due to
the tidal instability, the efficiency of an X-ray production was progressively decreasing. Secondly,
during stage A the X-ray flux was very stable (and probably a bit higher than at the end of the initial
rapid decline), but at the time of the stage A--B transition the flux began declining again and in
1--2 days dropped notably ($>$30\%) to lower values. Thus, although the case of GW~Lib is not as striking
as in \SSS, it also supports our hypothesis that the transitions between the superoutburst stages are
due to transformations in the entire accretion disc.

This conclusion is not obvious from prior theory and observations. It is widely accepted that particle
orbits in the inner accretion disc are approximately circular and that the tidal influence of the secondary
becomes significant only in the outer disc, where the tidal dissipation dominates the viscous shear
dissipation \citep{MurraySPH}. As discussed by \citet{OsakiMeyer03}, the tidal dissipation of superhumps
occurs in the outer 10 percent of the disc. However, \citet{OsakiKato13} proposed that the decrease of the
superhump period during the stage A--B transition can be understood as being due to the propagation of the
eccentricity wave to the inner part of the disc. As a result, the overall precession rate of the disc will
be determined by its larger portion. Although \citet{OsakiKato13} admitted that their interpretation is still
a matter of speculation, our results indicate that the changes in the disc structure during the superoutburst
stage transitions penetrate deeply into the inner, X-ray emitting, regions.

\subsection{Post-outburst decline}

The long-lasting post-outburst decline of the X-ray flux, which was exhibited by \SSS, indicates
strongly that the enhanced accretion onto the WD continued for several hundred days after the rapid
fading stage. However, such an enhanced accretion is expected to last only from the initial rise to an
outburst when the heating wave reaches the BL, to the time when the cooling wave does the same, transforming
the whole disc into the cool, quiescent state. According to the DIM \citep{Cannizzo93,Lasota2001}, the
viscosity in the quiescent accretion disc is low, especially in WZ~Sge-type stars \citep{Smak93,Lasota2001},
thus the viscous radial drift of matter is also very slow \citep{Cannizzo93}. Moreover, in the cool disc the
mass accretion rate is expected to decrease steeply with decreasing radius ($\dot{M}$ $\propto$ $r^{3}$ in such
a disc, see e.g. \citealt{Cannizzo93,Ludwig94,Menou99}), thus almost no matter should reach the inner disc
and the X-ray flux is expected to be low and stable. Nevertheless, despite the transition of the accretion
disc of \SSS\ to its quiescent state soon after the rapid fading stage (this is evident e.g. from the
replacement of the broad absorption troughs in the optical spectra by strong double-peaked emission lines,
usually observed in quiescent accretion discs -- \citetalias{Paper1}),
the X-ray flux declined for at least
$\sim$500~d, while the optical flux certainly declined continuously for $\sim$500 d.
Such long post-outburst declines are possibly a distinguishable property of the WZ~Sge-type
stars\footnote{Long optical declines are routinely observed in the WZ~Sge-type stars \citep{KatoWZ}.
As for the X-ray declines, besides \SSS\ only two other WZ~Sge-type stars were observed in X-rays at this stage
-- GW~Lib and WZ~Sge itself. Although those observations were sparse, both objects showed a significant
decrease of the X-ray flux between the beginning of the post-superoutburst decline and the quiescent state
\citep[see also Fig.~\ref{Fig:GW_Lib_LC}]{GWLibXrays,WZSgeQ}.}, whereas ordinary SU~UMa-type stars do not show
them \citep[see e.g.][]{OsakiKato13}.

It is particularly interesting that although X-rays and the optical flux predominantly originate at different
sites -- X-rays close to the WD, whereas the optical flux comes from the outer parts of the disc --
their post-superoutburst decline in \SSS\ was quite similar (Fig.~\ref{Fig:DeclineLog}). In Fig.~\ref{Fig:X_Opt}
we show the optical/X-ray flux diagram which includes all the post-superoutburst data and some representative
data-points from the rise and late plateau stages. The most prominent features of this diagram are

\begin{enumerate}
 \item A nearly linear dependence between the X-ray and optical fluxes during the post-superoutburst
 decline.

 \item The dynamic range of the X-ray flux in the post-outburst decline is less than that of the optical flux:
 \begin{equation*}
    R\equiv\frac{F_{x,max}}{F_{x,min}} / \frac{F_{opt,max}}{F_{opt,min}}\approx 0.6\,
 \end{equation*}
 where $F$ are the maximal and minimal X-ray and optical fluxes.

 \item The fitted line in the optical/X-ray flux diagram crosses the zero optical flux axis with a
 non-zero X-ray flux. This indicates that the dependence between the X-ray and optical fluxes must
 steepen significantly toward very low fluxes and accordingly very low mass accretion rates.
 This hypothesis is confirmed by the observations in quiescence, during which the X-ray flux
 is found to be significantly lower than that expected from the linear fit.

\end{enumerate}

It is interesting to note that features 2) and 3) resemble strongly the behaviour of two outbursting, short
orbital-period intermediate polars \object{CC~Scl} and \object{FS~Aur} in both outbursts\footnote{In intermediate
polars (IPs) the accretion occurs onto the magnetic poles of the WD. Because there is no BL in such a system,
the X-ray flux, even during an IP outburst, is not expected to be suppressed, as in most non-magnetic DNe,
but to increase proportionally with the mass accretion rate onto the WD.} and in quiescence \citep{NeustroevIP}.
In particular, the ratio of the outburst amplitudes $R$ in X-rays and in the optical is also close to 0.6
in both objects.

A linear dependence between the X-ray and optical fluxes during the return to quiescence indicates
that matter did not pile up in the outer disc, as expected in low-viscosity quiescent discs, but still
drifted toward the WD relatively freely through the entire disc, more resembling the behaviour of a
high-viscosity steady-state disc.

\section{Possible scenarios explaining the raised viscosity of a post-superoutburst disc}

There is no easy explanation as to why the viscosity remained elevated during the long post-outburst decline
to quiescence, when the disc has already cooled down below the critical temperature, associated with
the lower bend in the S-curve\footnote{This is evident from optical photometry and spectroscopy: during
a few days after the rapid fading from the superoutburst the brightness of \SSS\ has dropped by $\ga$4
mag; the spectra obtained on $T$=72 and later show very wide absorption troughs at the wings of the Balmer
emission lines; these troughs are formed in the WD photosphere, indicating a very low accretion luminosity
(see Section 4.1 in \citetalias{Paper1}).}, and why the viscosity finally switched to a low value, allowing
\SSS\ and other WZ~Sge-type stars to reach quiescence. A possible mechanism is the heating of the inner
disc by the hot WD. Such a scenario has originally been proposed by \citet{King97} in order to solve
the so-called UV delay problem.



\citet{King97} showed that a WD with an effective temperature high enough (a few
$\times$~10$^{4}$~K) should heat the inner disc up to the transition radius $R_{\rm tr}$ of a few
$\times$~$R_{\rm wd}$ and maintain it in the hot, high viscosity state.
\citet{King97} argued that at the end of an outburst the cooling wave
cannot propagate within $R_{\rm tr}$, where the disc remains hot and matter accretes to the WD much faster
than at the cool region outside $R_{\rm tr}$, resulting thus in a very low surface density of the inner disc.
In this respect we note that although the inner hot flow structure and the exact mechanism of X-ray production
in this model are still uncertain, the emitted X-ray flux should be proportional to the mass accretion rate at the
radius $R_{\rm tr}$, which in turn depends on the temperature of the WD: the hotter WD is, the higher X-ray flux
is expected. An important baseline for our discussion is that the WD temperature in DNe is not constant, but is
increasing during outbursts up to about 30000~K and then slowly decreasing in quiescence \citep{Sion95,Gansicke96}.
The latter should result in shrinking $R_{\rm tr}$ and accordingly in decreasing the X-ray flux.

The WD heating
during outbursts is primarily compressional heating by the additional amount of mass added on the WD envelope.
The thermal response of a WD to compressional heating has been widely studied in the literature. From a theoretical
point of view, the longer the outburst, the more mass is added to the envelope, the deeper the WD is heated
and so the longer the post-outburst cooling \citep{Sion95,Piro2005}. This is confirmed by observations:
the temperature decay time is indeed significantly longer after a superoutburst than after a normal outburst.
In SU~UMa-type stars, it is days after a normal outburst, weeks after a superoutburst \citep{Gansicke96,Long2009}.
For WZ~Sge-type stars, the superoutbursts are much longer, they dump much more mass and the WD cooling times are
years, 1--2 orders of magnitude longer than in ordinary SU~UMa-type stars \citep{Sion99,Godon2006,Toloza2016}.
These time-scales are consistent with those of post-outburst declines, and, qualitatively, the proposed scenario
appears to be promising. Quantitatively, both the X-ray flux in GW~Lib (Fig.~\ref{Fig:GW_Lib_LC})
and its WD temperature \citep{Bullock2011,Toloza2016,Szkody2016} are continuing to decline.
This indicates that the transition radius $R_{\rm tr}$ has not yet reached the WD surface.

Unfortunately, the proposed scenario is incomplete in the sense that it fails to explain the extended post-outburst
decline of the optical flux. The optical light originates beyond $R_{\rm tr}$, where the disc is already cool and
is not expected to cool down any more. Perhaps, it can be explained by strong boundary effects between the hot and
cold parts of the disc. At the moment, it is still unclear how exactly the transition from an inner hot disc to an
outer cold disc occurs. \citet{StehleKing} suggested that this transition region is extended and semistable.
Another problem standing against the proposed model is that the long post-outburst declines were observed not only
in the WZ~Sge-type stars, but also in low mass X-ray binaries (LMXBs), in which the primary, accreting component is
a neutron star or a black hole. For example, the black-hole candidate \object{GRO~J0422+32} after the main outburst
in 1992 exhibited several rebrightenings and a long decay, similar in appearance to those observed in WZ~Sge-type
stars \citep{Kuulkers98}. Obviously, the WD heating cannot explain such a behaviour of a black-hole binary.


Attempting to find an adequate mechanism that is able to maintain the post-superoutburst disc viscosity
higher than that at the pre-superoutburst level, when the system was in proper quiescence, we paid
attention to the fact that many WZ~Sge-type stars exhibit so called ``late superhumps'' for months
following superoutbursts \citep{KatoWZ}. In \SSS\ we traced superhumps through
the complete decline stage until at least 420~d after the rapid fading, but were unable to detect them
in quiescence \citepalias{Paper1}. We suspect that the disappearance of superhumps in quiescence is
not just a coincidence, but instead the superhump disappearance stops the decline trend. In other words,
we propose that an elliptical distortion of the disc can somehow stimulate a matter drift through the disc.

The exact mechanism responsible for this is not clear. \citet{Osaki97} proposed that a temporal
enhancement of viscosity in the quiescent disc could be produced by its highly turbulent state just
after the superoutburst. They suggested that a possible cause of a strong turbulence in the disc after
the superoutburst can be the tidal instability. Indeed, in
Section~\ref{Sec:Xsuperhumps} we presented evidence that the inner disc regions are probably under the
influence of disc precession. Thus it is plausible that particles on quasi-elliptical orbits can penetrate
the inner disc. The presence of a viscosity, even in an extreme case of WZ~Sge-type systems, causes
the orbits to intersect, leading to efficient gas heating and keeping thus the inner disc in a hot state
with an increased viscosity. While the elliptical disc is slowly shrinking after the superoutburst, matter
flow to the inner disc is also decreasing, resulting in gradual cooling. However, when the disc shrinks
inside the 3:1 resonance radius, it suddenly returns to a near-circular form and finally reaches the
quiescent state with a very low viscosity.

\section{Summary and concluding remarks}
\label{Sec:Summary}

We have analysed extensive X-ray observations of the WZ~Sge-type DN \SSS\ during its superoutburst in
2013, decline and subsequent quiescence. We collected 60 {\it Swift}-XRT observations of \SSS\ between
2013 January 6 and 2013 July~1. Four follow-up observations were performed in 2014, 2015, 2016 and 2017.
The total exposure time of our observations is 86.6~ks. The results obtained were compared with the
properties of another WZ~Sge-type DN, GW~Lib, for which new X-ray observations were also obtained. Our
primary findings can be summarised as follows:

\begin{enumerate}
 \item \SSS\ showed a 5 times higher hard X-ray luminosity during its superoutburst than during its
       subsequent quiescence.
 \item We detected a sudden X-ray flux change in the middle of the superoutburst coincident with a
       change in superhump behaviour. A similar X-ray behaviour was also detected in GW~Lib.
 \item The post-outburst decline of the X-ray flux of \SSS\ lasted for at least 
       $\sim$500~d, while the optical flux certainly declined continuously for $\sim$500~d.
 \item We found no direct evidence of the expected optically thick BL in the system during the outburst.
       If it is present at $L$$\sim$2\tim{34} \ergPerS\ then its blackbody temperature cannot be
       significantly different from kT$\sim$15 eV.
 \item The rapid decline of the X-ray flux of GW Lib, which was observed during the first 10 days after
       the initial rise at the beginning of the superoutburst, ended exactly at the time of
       the first appearance of superhumps. This stage of the superoutburst of \SSS\ was missed.
 \item New observations of GW~Lib in 2017 showed a decrease of X-ray flux in comparison to that observed
       after the optical outburst eight and nine years before. However, 10 years after the superoutburst,
       the X-ray flux is still about five times larger than was measured before the superoutburst.
\end{enumerate}

We have showed for the first time that the X-ray flux from two WZ~Sge systems is linked to their simultaneous
superhump behaviour, thus linking the inner disc properties to those of the outer (possible whole) disc. Indeed,
superhumps are caused by the disc precession, and their appearance, evolution and stage transitions reflect
geometrical and/or dynamical transformations of the asymmetrical disc. This result suggests that models
for accretion discs in high mass ratio accreting binaries are currently incomplete. The very long decline
to X-ray quiescence is also in strong contrast to the expectation of low viscosity in the disc after outburst.
We propose that the disc
precession is a cause of an enhanced viscosity and an increased temperature of the inner disc, stimulating
thus a matter drift through the disc during the decline stage of superoutbursts. These unexpected findings
can have important implications beyond CVs, for example to LMXBs, in which a precessing accretion
disc can be formed during outbursts \citep{Smith07}.

\begin{acknowledgements}
We are thankful to Neil Gehrels for the support showed us during this and other projects. May he rest
in peace. We thank the \textit{Swift} team for executing the observations. We thank Boris G{\"a}nsicke
for useful discussion of the white dwarf cooling problem. We are thankful to the anonymous referee whose
comments helped greatly to improve the paper. This research has made use of data obtained
through the High Energy Astrophysics Science Archive Research Center Online Service, provided by the
NASA/Goddard Space Flight Center. This work made use of data supplied by the UK Swift Science Data
Centre at the University of Leicester. We acknowledge with thanks the variable star observations from
the AAVSO International Database contributed by observers worldwide and used in this research.
KLP, JPO and APB acknowledge the support of the UK Space Agency. TRM is supported by the STFC under grant
ST/L000733. The work was supported by the Deutsche Forschungs-Gemeinschaft (DFG) grant WE 1312/51-1 and
the Russian Foundation for Basic Research grant 16-02-01145-a (VFS). The work of VFS was also funded by
the subsidy allocated to Kazan Federal University for the state assignment in the sphere of scientific
activities (3.9780.2017/8.9). SVZ acknowledges PAPIIT grant IN-100617 for resources provided towards this
research.
\end{acknowledgements}

\bibliographystyle{aa} 
\bibliography{SSSx.bib} 
\listofobjects
\end{document}